\documentclass[journal]{IEEEtran}
\usepackage{graphicx}
\begin{document}
\title{Powering via Cooling Pipes:\\ an Optimized Design for an SLHC Silicon Tracker}
%
%

\author{Wim de Boer and 
        Jochen Ebert
\thanks{Manuscript received November 14, 2008. }
\thanks{W. de Boer and J. Ebert are with the University of Karlsruhe, Postfach 6980, 76131 Karlsruhe, Germany (telephone: +49-721-608-3593, e-mail: deboer@physik.uni-karlsruhe.de). }}%

\maketitle
\pagestyle{empty}
\thispagestyle{empty}

\begin{abstract}
Silicon trackers at the SLHC will suffer high radiation damage from particles produced during the collisions, which leads to high leakage currents. Reducing these currents in the sensors requires efficient cooling to  -30 $^0$C. The large heat of evaporation of CO$_2$ and the low viscosity allows for a two-phase cooling system with thin and long cooling pipes, because the small flow of liquid needed leads to negligible temperature drops. In order to reduce the material budget a system is proposed in which a large scale tracker requiring ca. 50 kW of power is powered via 1-2 mm diameter aluminum cooling pipes with a length of several m. These long cooling pipes allow to have all service connections outside the tracking volume, thus reducing the material budget significantly. The whole system is designed to have negligible thermal stresses. A CO$_2$ blow system has been designed and first tests show the feasibility of a barrel detector with long ladders and disks at small radii  leading to an optimized design with respect to material budget and simplicity in construction.
\end{abstract}


\section{Introduction}
%
%
%
%
\IEEEPARstart{T}{he} Large Hadron Collider (LHC) collider at CERN is expected to start data taking in 2009
with
the luminosity  ramp up to the design goal of 10$^{34}$/cm$^2$/s over the next few years.
After 2016 a proposal to extend
the physics potential of the LHC with a major luminosity upgrade to the Super LHC (SLHC) has been endorsed
by the CERN council strategy group \cite{slhc}. The goal is a factor ten increase in luminosity.
This imposes severe requirements on the silicon tracking devices, which need to be completely
redesigned in order to cope with the order of magnitude increase in occupancy and radiation damage.
The increase in occupancy requires a corresponding increase in granularity, while the radiation
hardness can be improved by a different sensor design combined with cooling to lower temperatures in
order to reduce the leakage current.

Radiation damage to silicon sensors creates defects with  energies between the valence and conduction band. This increases both the leakage current at a given temperature and the depletion voltage. If one uses n-strips on p-type wafers the depletion will start from the n-strips, so there will be no undepleted layer between the sensor and the strips.  Furthermore, it was shown that even after 10$^{16}$ n/cm$^2$, which is the fluence expected at the SLHC at the innermost layers during its lifetime, the collected charge is still around 7000 electrons, which is roughly one quarter of the non-irradiated sensor \cite{casse}.

 If one reduces the noise both, by cooling the sensor and by reducing the sensor capacitance (by reducing the strip length) it seems feasible to get a similar signal/noise ratio as for non-irradiated samples. Shorter strips ("strixels") are anyway required to reduce the channel occupancy for the high luminosity of the SLHC.

Another requirement is the reduction of the material budget in order to reduce the interactions in the tracker, which lead to a distortion of the momentum measurements and an increase of the multiplicity by photon conversions. One expects up to 400 interactions per bunch crossing at the SLHC with typically 1000 charged tracks per bunch crossing. If most photons convert into electron-positron pairs, as they will do with the present material budget, the charged multiplicity will be doubled to tripled.



\begin{figure}[!t]
\centering
\includegraphics[width=3.5in]{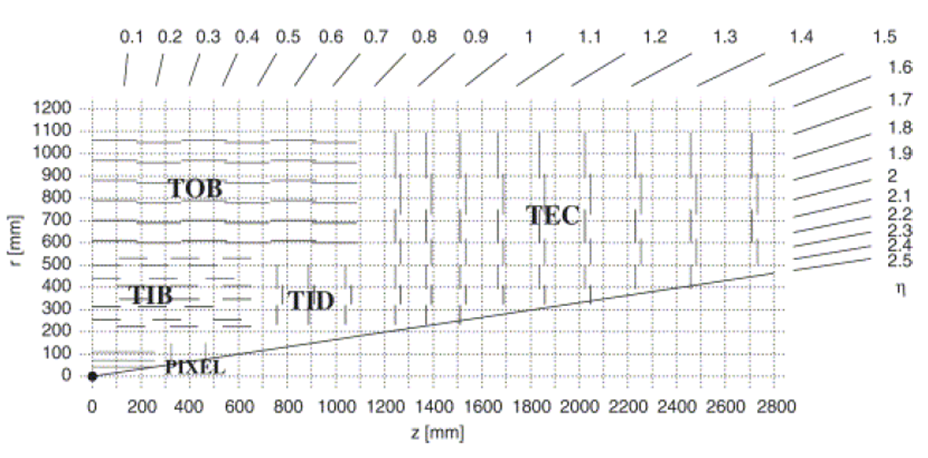}
\caption{The subdetectors of the CMS silicon tracker system: TOB=outer barrel, TIB=inner barrel, TID=inner disc, TEC=endcaps, PIXEL=pixeldetector. The pseudo-rapidity values $\eta$ have been indicated as well. }
\label{f1}
\end{figure}
\begin{figure}[!t]
\centering
\includegraphics[width=3.5in,height=3.5in]{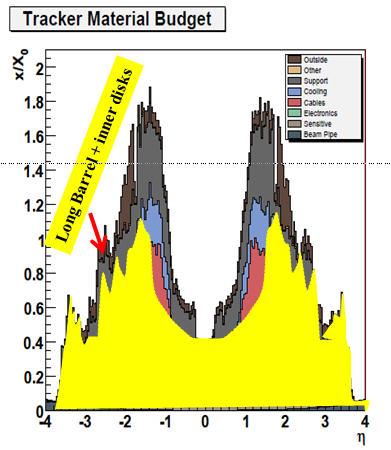}
\caption{The material budget in the present CMS tracker. A large fraction for $1.2<\eta<2.2$ originates from the service connections for cooling and power to the inner barrel and disk detectors, which pass in front of the endcap detectors \cite{cms}. This can be avoided by having a long barrel (see Fig. \ref{f3}) with all service connections outside the tracking volume. This leads to a material budget indicated by the light shaded (yellow) area.}
\label{f2}
\end{figure}
Here we present a tracker design based on CO$_2$ two-phase cooling. CO$_2$ allows for long cooling pipes with still negligible temperature gradients, because of the small pressure drop (owing to the large latent heat and correspondingly small mass flow and the low viscosity). CO$_2$ cooling is efficient between 20 and -50 $^0$C, so the leakage currents can be made negligible, while the long cooling pipes allow  the use of  3 m long ladder type detectors, thus paving the way for having all service connections outside the tracking volume. By combining the functions of cooling pipe, mechanical support and current leads the material budget can be additionally minimized. First tests with a simple CO$_2$ blow system prove the feasibility of the approach.
\section{Strawman design}
\begin{figure}[!t]
\centering
\includegraphics[width=3.5in]{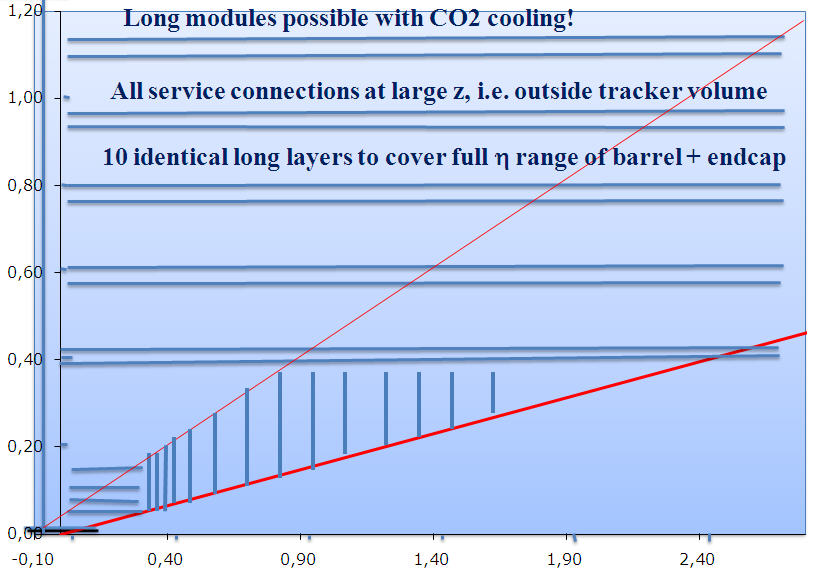}
\caption{A strawman design of a possible SLHC detector with a long barrel and endcaps at small radii. The units on the axis are in m. This design allows to have all service connections outside the tracking volume, but requires efficient and long cooling pipes, which is possible with a CO$_2$ two-phase cooling system. }
\label{f3}
\end{figure}

\begin{figure}[!t]
\centering
\includegraphics[width=3.5in]{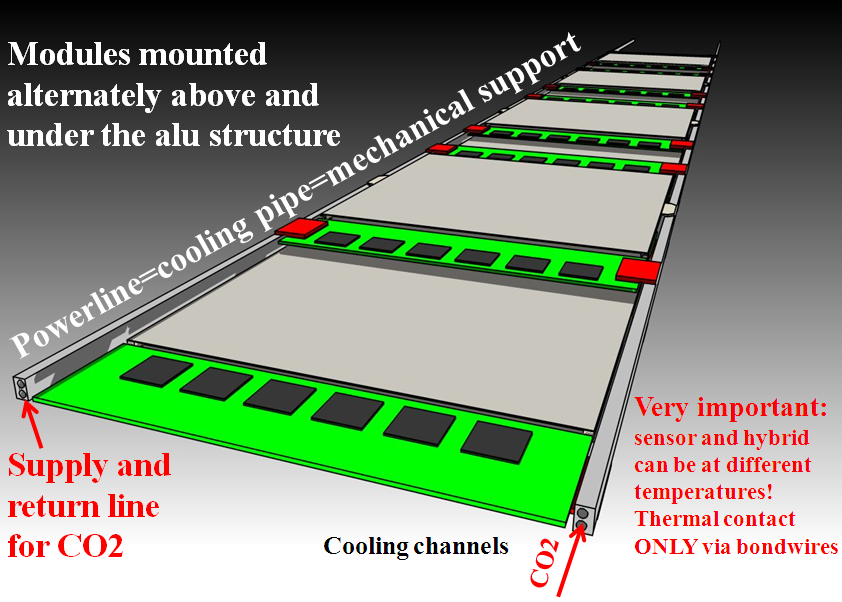}
\caption{A schematic view of the principle of powering via cooling pipes with the sensors (grey) and hybrids (green with black front end chips) mounted separately on the power lines. The current is transferred to the hybrids via the  cooling blocks (red), which provide simultaneously the mechanical support. The whole structure is mounted on a 5 mm thick Rohacell foam plate, which forms with the  cooling pipes fixed between the cooling blocks a stable sandwich structure.}
\label{f4}
\end{figure}

A traditional layout of a silicon tracker has a barrel with endcaps and a pixelated detector near the beam pipe, as shown in Fig. \ref{f1} using the 200 m$^2$ silicon tracker of CMS as an example \cite{cms}. The main reason for the endcaps is that in the forward direction the particles traverse  horizontal detectors under a small angle $\theta$,    thus increasing the material thickness by a factor $d/\sin\theta$.    Therefore vertical detectors are  much better with respect to the material budget for small angles $\theta$, since in this case the traversed material is $d/\cos\theta$.    However, in this case the cooling pipes, current leads and endflanges of the horizontal detectors  are just in front of the endcaps, which increases the material budget in the forward region by as much as 0.8 $X_0$, as shown in Fig. \ref{f2} for the present CMS Silicon tracker.
To avoid this would require long barrel detectors, which seemed difficult to cool. However, with CO$_2$ two-phase cooling this is  possible and therefore it seems better to use only horizontal detectors, as shown in Fig. \ref{f3}.  In this case all service connections and interconnect boards for the read out can be at the end of the barrel, i.e. outside the tracking volume. Also optocouplers on the interconnect boards would be further away from the center, which is important, since these are known to  suffer severe radiation damage after a fluence of $10^{14} n/cm^2$. Furthermore, they consume a significant amount of power, which need not to be entered into the tracking volume. All signals from the hybrids have then to be transferred to the interconnect board, e.g. by LVDS signals via aluminized kapton cables. An additional reduction of the material budget can be obtained by using the cooling pipes as power lines, as sketched in Fig. \ref{f4}. Details will be discussed in the next section.

\begin{figure}[!t]
\centering
\includegraphics[width=3.5in]{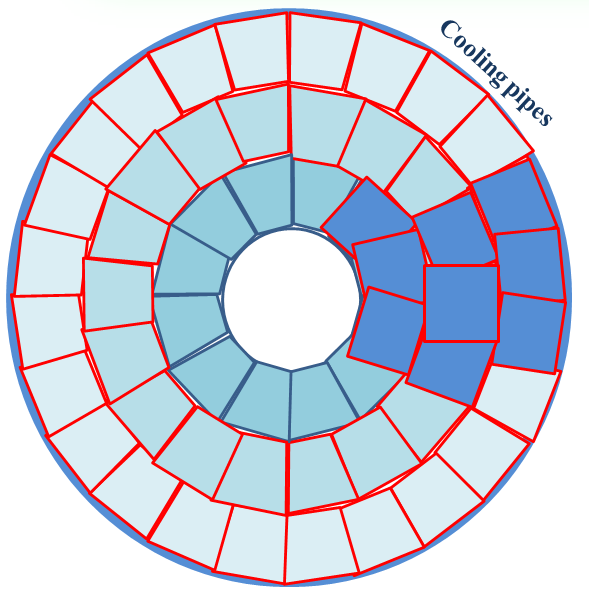}
\caption{A layout of the endcap detectors at small radii. The dark shaded sensors are rectangular instead of trapezoidal sensors, which have the advantage of being the same ones as in the barrel, but this results in a somewhat higher overlap.}
\label{f5}
\end{figure}

To get enough tracking points for tracks in the forward direction requires then to have disks at small radii. A disk layer for radii between 5 and 35 cm is shown in Fig. \ref{f5}. The rings can be constructed in the same way as the ladders of the barrel, i.e. with cooling pipes at each side which act at the same time as power lines. The only difference is that the cooling pipes are now bent in half circles with the connections on the inside, i.e. outside the tracking volume. Half circular disks are are needed to be able to install or replace the detectors with the beampipe installed. The four inner horizontal layers and four vertical disks are pixelated, most likely similar to the present pixel detectors.  Note that  the pixel detectors as well as the strixel disks can be exchanged without moving the major part of the detector, the barrel detectors. This is important, since for these detectors the radiation damage is much higher than for the barrel, so they may need replacement.

 Having all service connections outside the tracking volume reduces the material budget  by 40\% or more, as shown by the light shaded (yellow) area in Fig. \ref{f2}.
Note that in the center the total material budget is similar, around 0.4  X$_0$, as expected, since in this area there are no service connections and the number of layers is similar.

The luminosity at the SLHC will be an order of magnitude higher than at the LHC. Requiring the same occupancy implies that the number of channels increases also by a factor between 5 to 10, e.g. by having pixel detectors or shorter strips ("strixels"). The power will not increase by an order of magnitude, since the
smaller feature size of today's and future electronics is expected to reduce the power per channel by a factor 5 or more\cite{raymond}. So the total power for the CMS upgrade is expected to be of the order of 35 kW, similar to the present CMS tracker.

If one opts for a long barrel of identical ladders, one needs only one type of sensors with eventually different pitches and strixel lengths for the inner and outer ladders in order to reduce the number of channels. This simplifies the detector construction enormously in comparison with the present detector. Long ladders require efficient cooling. We first discuss  the module design integrated on a long ladder with cooling pipes as power lines integrated in the same mechanical structure and then the cooling system.
 \section{Ladder design}
\begin{figure}[!t]
\centering
\includegraphics[width=3in]{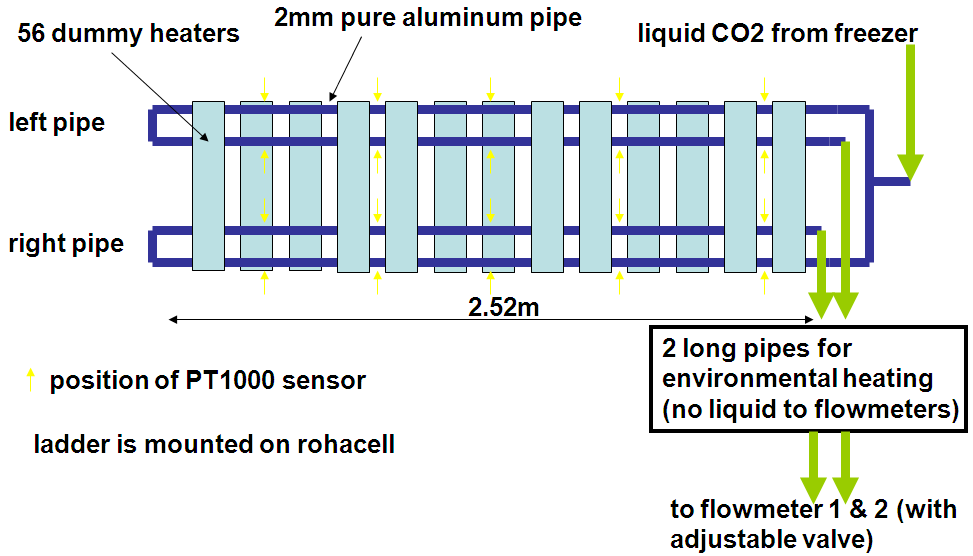}
\includegraphics[width=3in]{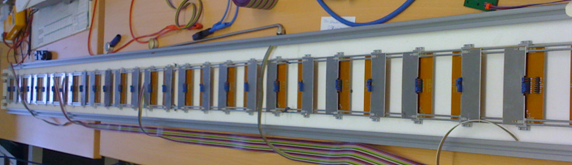}\\[2mm]
\includegraphics[width=1.5in]{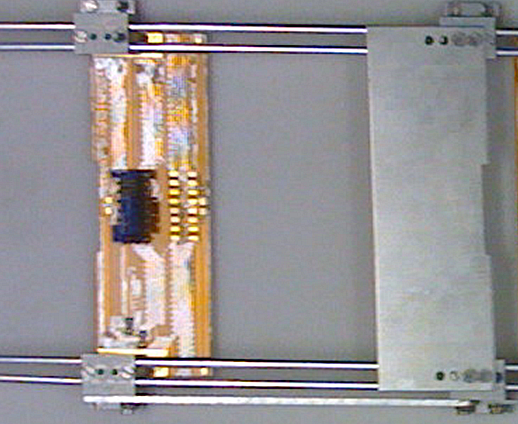}
\caption{A schematic view of a ladder (top), a photograph of a ladder (middle) and detail of a hybrid mounted on the cooling tubes. All 56 modules in a single ladder are powered in parallel. The hybrids are mounted alternating above and below the cooling tube to allow overlap between the sensors. To transfer the heat from the hybrid to the cooling tubes the kapton must be glued on a carbon fiber or Al frame. The latter was used for the moment. The thin Al plate can be seen at the bottom right.  }
\label{f6}
\end{figure}

\begin{figure}[!t]
\centering
\includegraphics[width=3in]{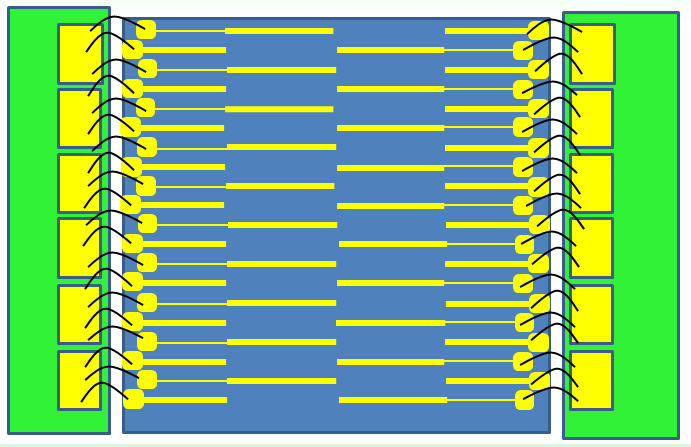}
\caption{A schematic view of the sensor with short strixels and bond pads at each side of the sensor.}
\label{f7}
\end{figure}

\begin{figure}[!t]
\centering
\includegraphics[width=3.5in]{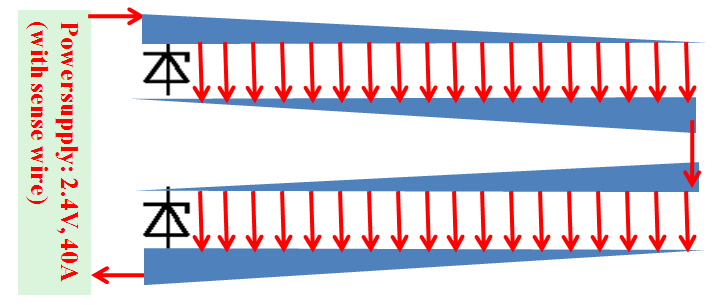}
\caption{A schematic view of the current flow in two neighboring ladders. All 56 hybrids in a single ladder are powered in parallel, while the two ladders are powered serially. The thickness of the power lines varies in order to have the same voltage drop on the positive and negative line, thus providing a constant voltage  to each sensor. In practice this can be realized by an additional Al sidebar, as shown at the bottom of Fig. \ref{f6}.}
\label{f8}
\end{figure}

 As mentioned in the introduction, the material budget can be reduced by long ladders having all service connections outside the tracking volume as well as  combining the mechanical support, the cooling pipe, which acts simultaneously as power line, into a single mechanical structure. A possible design is sketched in Fig. \ref{f4}.
 The pair of cooling pipes on each side of the detector (inlet and outlet) are most easily realized by extruded Al tubes. Pure Al, which is easy to extrude, has the additional advantage that the electrical resistivity drops by about 30\%, if one cools the tube from  +40 to -40 $^0$C. For a 2.5 m long pair of cooling pipes with an ID of 1.5 mm and outer OD of 3 mm the total resistance is about 5 m$\Omega$, so they form excellent power lines. In order to avoid a contact resistance to the hybrid, the Al tubes have to be chromatized, which is  a standard process.
 The sensors are glued to a piece of Al,  shown in red, which serves simultaneously as power connector, the mechanical support and the cooling point. The hybrids and the sensors are both screwed on the cooling pipes, but there is no further mechanical or thermal contact between them, so the hybrid and sensor are cooled independently and thus can operate at different temperatures. Thermal stresses are reduced, since the hybrid and sensor move together with the cooling pipes during cooldown.  For the tests the hybrids consisted of a simple kapton printed circuit with a resistive line between the power pads  and a PT100 thermometer, which can be read out via an USB interface. The power pads of copper on the kapton are pressed against the cooling block by the mounting screw, which forms simultaneously the power connector. A picture is shown in Fig. \ref{f6}. In total  56 hybrids are screwed  between the cooling pipes.  Of course, in such a scheme all modules are powered in parallel, which raises several questions. First, what is the voltage drop along the power line.
 The current needed for a single ladder is estimated as follows. Each sensor of 9x10 cm is assumed to have strixels of 2.2 cm with bond pads at the outer edge, as indicated in Fig. \ref{f7}. Assuming a pitch of 130 $\mu m$ leads to a total of 12x256 channels, i.e. 6 front end chips with 256 channels  bonded to each side of the detector. If we assume new front ends need 0.5 mW per  channel \cite{raymond}, if the 0.13 $\mu m$ or even smaller feature sizes will be used and add 25\% for the control power, one needs 1.6 A at 1.2 V per sensor. For 56 hybrids this corresponds to about 50 A and 60 W per ladder. To get all connections at large z requires that the current is returned, e.g. via the cooling pipes of the neighboring  ladder, as sketched in Fig. \ref{f8}. In this case the current is used twice, thus reducing the total amount of current in the power lines between the power supplies and the detector. The total current needed for such a design is of the order of 12 kA (33 kW) for a total of 64 million channels. This is to be compared with 15 kA and 33 kW in the present CMS detector.

 If the input current flows in both power lines in the same direction by returning the current in the neighboring ladder, one has the same voltage drop on each power line, so one naively expects no influence
 of the voltage drop on the hybrids, since each hybrid would see the same voltage difference between the power lines. However, at the beginning of the ladder the positive power line carries 50 A and the negative only the current from the first hybrid, i.e. around 1 A, so the voltage drops are not equal in both power lines. This can be improved by increasing the thickness of the positive power line at the beginning, as  shown at the bottom of Fig. \ref{f6} by the additional piece of Al between the power connectors (=cooling points). In the middle of the ladder both power lines carry a similar current, so here nothing needs to be done, but at the end the returning power line needs to have additional material, since here  the full current flows. By this voltage compensation in the power lines all hybrids can have the same voltage with a precision better than 20 mV.
   The hybrids have not exactly a common ground because of the voltage drop of about 0.1 V along the power lines, so the control and signal lines better be differential LVDS type lines with a standard AC coupling to the interconnect boards at the end of the ladder. Alternatively,  a differential LVDS receiver with a somewhat higher source voltage can easily cope with common mode offsets of 100 mV. 

The detectors have been mounted on a
  5 mm thick support plate made from Rohacell foam, a strong lightweight material, which has been used before as support in silicon detectors \cite{phenix}.  With the attached cooling structure this forms a  stable double layer sandwich structure which can easily take up the magnetic forces of the power lines, if the ladder is inside a solenoid. The radiation length of Rohacell is above 5 m, so it does not contribute significantly to the material budget. However, it is hygroscopic. Water absorption can be prevented by a  parylene coating \cite{phenix}. Alternatively, the support structure can be made out of carbon fiber.
\section{Cooling system}
\begin{figure}[!t]
\centering
\includegraphics[width=3.5in]{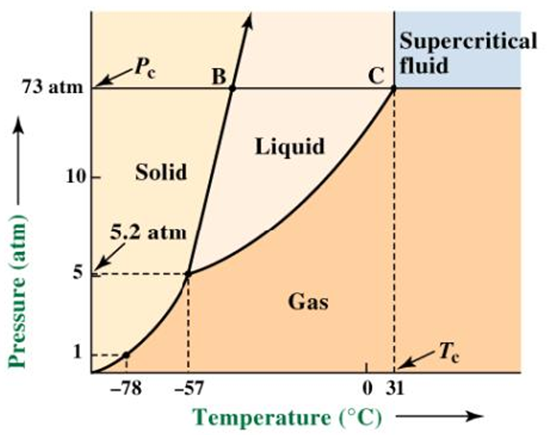}
\caption{The phase diagram of CO$_2$ showing that a two-phase liquid-gas system exists between -57 and 31 $^0$C, so this is the range of cooling liquid temperatures with vapor pressures between 5 and 73 bar. }
\label{f9}
\end{figure}

 The leakage current of heavily irradiated sensor can be reduced sufficiently by using sensor temperatures below -25 $^0$C.
 The high cooling power of CO$_2$ makes it a natural choice as refrigerant, especially since CO$_2$ becomes increasingly popular in replacing the climate unfriendly freons. This leads to many off-the-shelf components, like CO$_2$ liquid pumps. The temperature range for CO$_2$ 2-phase cooling can be read off from the pressure-temperature diagram in Fig. \ref{f9}: -57 to +31 $^0$C for pressures between 5.2 and 73 bar, respectively. This implies that during normal operations the pressure is below 20 bar, but after warm up to room temperature the pressure increases to around 50 bar at 20 $^0$C.

\begin{figure}[!t]
\centering
\includegraphics[width=3.5in]{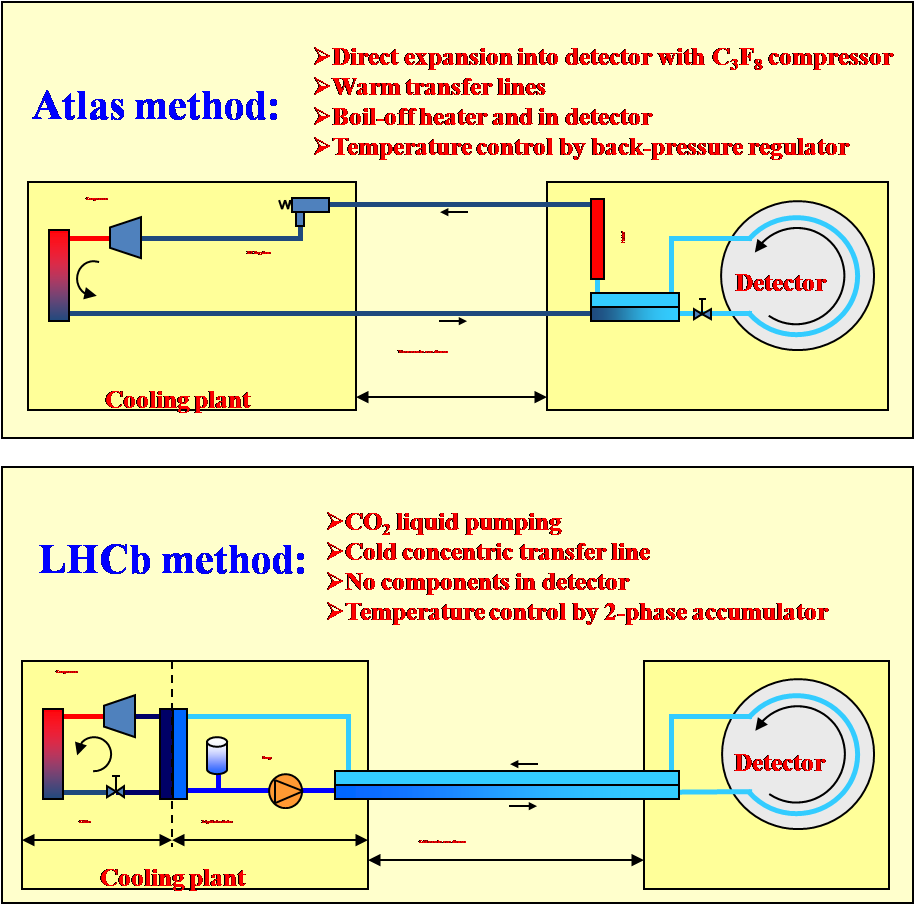}
\caption{Examples of two-phase cooling systems used at the LHC. Top: the ATLAS cooling system using a low pressure gas compressor for circulation, which requires heaters in the detector to evaporate all liquid, even in case no power is generated in the detector. Bottom: the LHCb cooling system, which pumps only liquid around using isolated in- and outlets to the detector, thus avoiding all active elements (switches, heaters) inside the detector. The temperature is regulated by the pressure in the accumulator. From B. Verlaat, NIKHEF, Amsterdam.}
\label{f10}
\end{figure}
What makes CO$_2$ so interesting for low mass cooling systems is its large enthalpy, which allows to cool close to 300 J for each g of CO$_2$ evaporated. This is to be compared with non-evaporative systems, like CMS, where only 5 J/g are obtained, if a temperature increase of 5 K is accepted. Therefore the flow rate of CO$_2$ liquid can be one to two orders lower compared to non-evaporative systems, thus reducing  the size of tubing, pumps etc. by a similar amount. In addition, the viscosity of CO$_2$ is low, so the pressure drop and the corresponding temperature drop in the cooling tubes is small. For evaporative systems one can either have a cooling plant with a gas compressor to liquify the evaporated gas or pump the liquid around with a liquid pump instead of using the compressor to build up pressure.  Both schemes have been used at the LHC, as shown in Fig. \ref{f10} for the ATLAS \cite{atlas} and LHCb \cite{lhcb} cooling systems. The ATLAS system is running with a low vapor pressure C3F8 coolant, while LHCb has opted for a high pressure CO2 cooling system, in which the liquid is pumped around.  Great features of the CO$_2$ cooling system are: i) no active elements like heaters or electric valves inside detector ii) standard liquid CO$_2$ pumps iii) standard primary chiller iv) the temperature of the whole system is controlled by only ONE  parameter, namely the vapor pressure in the accumulator, which can be increased by the heater and decreased by the chiller.
\begin{figure}[!t]
\centering
\includegraphics[width=3.5in]{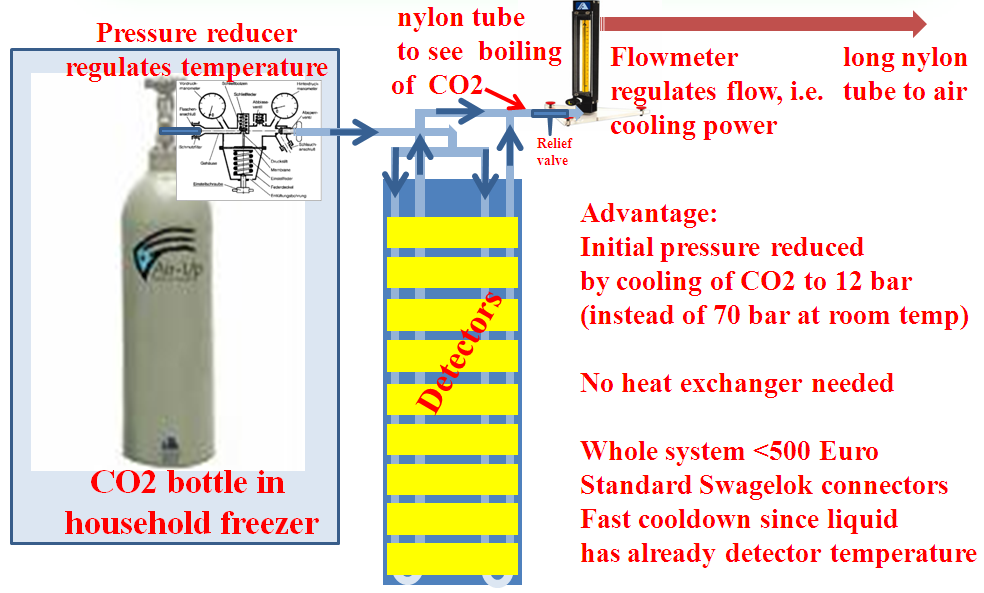}
\includegraphics[width=3.5in]{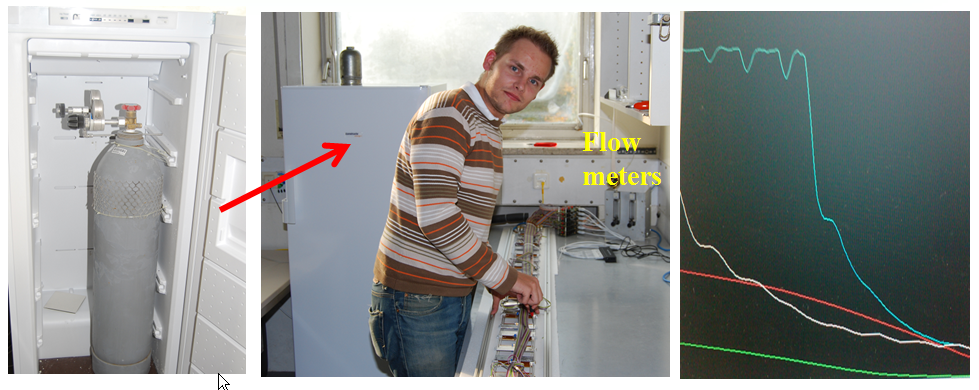}
\caption{Top: schematic  of a simple CO$_2$ blow system. The CO$_2$ bottle is precooled to -40 $^0$C in a household freezer. The CO$_2$ liquid is sent through 2 mm ID  Al cooling pipes of  a 2.5 m ladder of hybrids, consisting of kapton with resistors and PT100 temperature sensors (see Fig. \ref{f6}). The cooling pipes act at the same time as power lines and mechanical support. Bottom: photographs of the  system. The freezer is shown on the left.  The temperatures of different sensors as function of time are shown at the right. The blue curve shows the temperature of a sensor just be cooled by the arriving liquid.}
\label{f11}
\end{figure}
 Another advantage of having a liquid pump is that the system works very well without heat load, e.g. when the detector is switched off. Liquid CO$_2$ membrane pumps are widely used in the food industry and elsewhere, so they are cheap and easy to obtain including sensors in case the membrane would be leaking \cite{pumps}. Large liquid pumps with capacities of around 100 m$^3$/h are available, which are e.g. used on off-shore gas platforms to pressurize the natural gas with CO$_2$. For the cooling power of an SLHC tracker of about 50 kW pumps with a capacity of about 1 m$^3$/h are enough. Of course, pumping the liquid requires thermally isolated  outlets, but this is easily accomplished, especially since the temperature gradients are not large.

 Since CO$_2$ is non-toxic and non-flammable one can easily design small "blow"-cooling-systems, in which the CO$_2$ is blown to the air. The main precautions, which have to be taken, is that CO$_2$ gas has a higher density than air, so it will collect at the bottom. Therefore it has to be  blown outside the window, if one wants to prevent the risk of not having enough oxygen in the lab.
 A simple blow-system is shown in Fig. \ref{f11}. The CO$_2$ bottle is precooled in a commercial house-hold freezer to -40 $^0$C, which reduces the pressure to around 10 bar. In this case nylon tubing with standard Swagelok connectors can be used to transport the liquid to and from the detector. The nylon tubing is already a good isolator, so only little additional isolation is required.  After the detector a simple ball flow meter, consisting of a little ball inside a flow tube with a needle valve at the entrance for regulating the flow rate, is installed. The environmental heat into the nylon tube is enough to evaporate the rest of liquid from the detector before entering the flow meter.

 The whole system is regulated by just two knobs: i) the pressure reducer regulates the temperature of the liquid-gas mixture coming out of the bottle, which has a riser tube sticking into the liquid at the bottom; ii) the cooling power is regulated by the needle valve in the ball flow meter.

\begin{figure}[!t]
\centering
\includegraphics[width=3.5in]{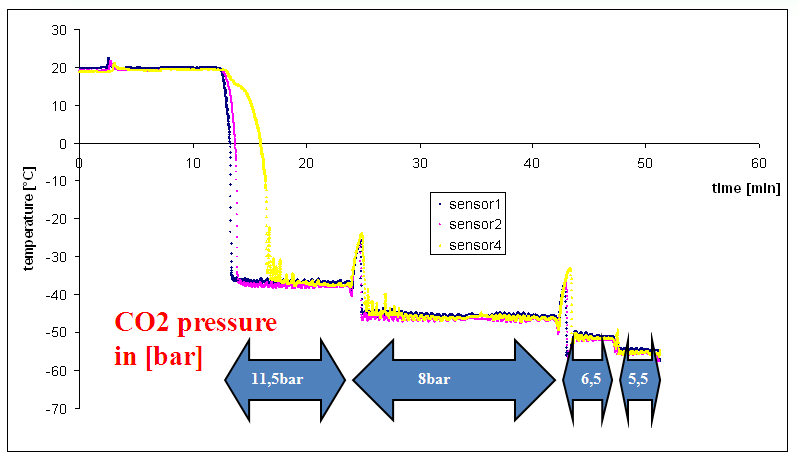}
\caption{The temperature response after changing the pressure with the pressure reducer mounted on the CO$_2$ bottle.}
\label{f12}
\end{figure}

\begin{figure}[!t]
\centering
\includegraphics[width=3.5in]{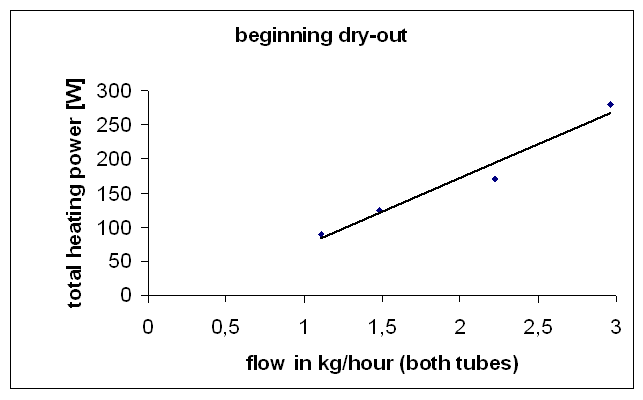}
\caption{The dissipated electrical power $Q_{elec}$ at a temperature of -31 $^0$C as function of the flow rate of CO$_2$ needed to prevent dry-out.
The total power $Q=Q_{elec} + Q_{envir}$ is proportional to the flow rate $\phi$, so the negative offset for $\phi=0$ represents environmental heat load $Q_{envir}$ of approximately 15 W. This is rather large, since the ladder was only inside a  box with a transparent cover to prevent condensation from the air, but this yields minimal isolation.}
\label{f13}
\end{figure}

  The long ladder, shown in Fig. \ref{f6}, with two pairs of 2x2.5 m long cooling tubes with an inner diameter (ID) of 2 mm and outer diameter (OD) of 3 mm has been tested. The temperature was set by the pressure reducer on the bottle, which kept the temperature on the whole ladder very well constant, as shown in Fig. \ref{f12}. Increasing the power in the hybrid resulted in evaporation of the liquid. If the liquid had evaporated before the end of the ladder, the temperature at the end of the ladder increased, i.e. dry-out occurred. Increasing the flow rate restored the cooling everywhere. Fig. \ref{f13} shows the power  against the flow rate required to be just above the dry-out condition. The minimum power estimated for a 2.5 m long ladder is 50 W, but clearly much higher powers can be sustained with 2 mm ID cooling tubes. The maximum power was not limited by the detector system, but by the maximum throughput of the flow meters.


\section{Conclusion}

The successful operation of a CO$_2$ cooling system in LHCb proves its feasibility \cite{lhcb}. This paves the way for a rather different design of a large radiation hard silicon tracker, since CO$_2$ cooling has two advantages: it allows to cool easily to temperatures below -30 $^0$, thus eliminating the leakage currents in a harse radiation environment and secondly, the high cooling power allows for tiny and long cooling tubes, which in turn allows to have long ladder type detectors with all service connections and interconnect boards outside the tracking volume.

 The number of layers for tracks at small angles with respect to the beampipe can be increased by a larger number of pixel layers and strixel disks at small radii. For pitches between 100 and 200 $\mu m$ and strixels around 2 cm the strixels can have their bonding pads all at the outside of a 9x10 cm sensor, so a similar module construction can be used as in present LHC detectors, except for having hybrids on each side of the detector. Although this increases the number of channels by a factor 5 to 10, the total power is expected to be similar because of the expected power reduction in future front end electronics with a smaller feature size. Serial powering of two ladders is proposed, which would keep the total current similar to the ones in present detectors, thus allowing to reuse the existing services and power supplies without the need for DC-DC converters.

  It is shown that  the material budget can be reduced by 40\% or more in the forward region of the CMS detector, if one adopts the strawman design of  long barrel detectors with inner disks at small radii (instead of the traditional endcaps), especially if one combines the functions of cooling pipes, power lines and mechanical support in a single structure. A  simple CO$_2$ blow system has been designed and first tests regarding powering via cooling pipes are encouraging. A final pair of ladders with real sensors and all interconnect boards at the end of a long ladder needs to be tested in order to check the stability of the proposed powering via cooling pipes and the transfer of signals to and from  the interconnect board at the end of the long ladder.

\section{Acknowledgment}
Special thanks go to Bart Verlaat for helpful discussions on CO$_2$ cooling. The project was supported by
the Bundesministerium f\"ur Bildung und Forschung under Grant 05 HS6VK1.

\end{document}